\documentclass[10pt,thmsa]{article}
\usepackage{amsfonts,amsmath,amsthm,amssymb}

\newcommand{\limfunc}[1]{\mathrm{#1}}
\newcommand{\func}[1]{\;\mathrm{#1}\;}
\newcommand{\qatopd}[2]{\genfrac{[}{]}{0pt}{}{#1}{#2}}

\newtheorem{lemma}{Lemma}[section]
\newtheorem{theorem}{Theorem}[section]
\newtheorem{corollary}{Corollary}[section]
\newtheorem{proposition}{Proposition}[section]
\newtheorem{remark}{Remark}[section]
\newtheorem{definition}{Definition}[section]

\begin{document}
\begin{titlepage}
\begin{flushright}
YITPSB-01-52 Preprint  \\
\end{flushright}
\vspace{0.5cm}
\begin{center}
{\Large {\bf Quantum superalgebras at roots of unity and
non-abelian symmetries of integrable models} }

\vspace{0.8cm}
{\large  Christian Korff$^1$ and Itzhak Roditi$^{1,2}$ }

\vspace{0.5cm}
{\em $^1$C.N. Yang Institute for Theoretical Physics\\
State University of New York at Stony Brook\\
Stony Brook, N.Y. 11794-3840, U.S.A.\\}
\vspace{0.2cm}
{\em $^2$Centro Brasileiro de Pesquisas F\'{\i}sicas\\ 
R. Dr. Xavier Sigaud 150\\ 
Rio de Janeiro, RJ 22290-180, Brasil}
\end{center}
\vspace{0.2cm}
 
\renewcommand{\thefootnote}{\arabic{footnote}}
\setcounter{footnote}{0}

\begin{abstract}
We consider integrable vertex models whose Boltzmann weights (R-matrices)
are trigonometric solutions to the graded Yang-Baxter equation. As is well
known the latter can be generically constructed from quantum affine
superalgebras $U_{q}(\hat g)$. These algebras do not form a symmetry algebra
of the model for generic values of the deformation parameter $q$ when
periodic boundary conditions are imposed. If $q$ is evaluated at a root of
unity we demonstrate that in certain commensurate sectors one can construct
non-abelian subalgebras which are translation invariant and commute with the
transfer matrix and therefore with all charges of the model. In the line of
argument we introduce the restricted quantum superalgebra $U^{\mathrm{\ res}%
}_q(\hat g)$ and investigate its root of unity limit. We prove several new
formulas involving supercommutators of arbitrary powers of the
Chevalley-Serre generators and derive higher order quantum Serre relations
as well as an analogue of Lustzig's quantum Frobenius theorem for
superalgebras.
\end{abstract}
\vfill{ \hspace*{-9mm}
\begin{tabular}{l}
\rule{6 cm}{0.05 mm}\\
korff@insti.physics.sunysb.edu\\
roditi@insti.physics.sunysb.edu\\
roditi@cbpf.br
\end{tabular}}
\end{titlepage}
\newpage



\section{Introduction}

The area of integrable models has proved to be one of the most fruitful
connections between physics and mathematics over the years. Especially the
study of the Yang-Baxter equation \cite{Yang,Baxter0} has lead to numerous
discoveries of new algebraic structures with a range of applications
reaching far beyond its origin in exactly solvable statistical mechanics
models \cite{BaxterB} and the quantum inverse scattering method \cite
{QISM,QISM1}. The widely known and best studied examples of such new
structures are quantum algebras (also called quantum groups) \cite
{Drin,Jimbo,FRT}. The latter are obtained from an affine Lie algebra $\hat{g}
$ as a particular $q$-deformation $U_{q}(\hat{g})$ and belong to the class
of non-cocommutative Hopf algebras. Once a coproduct is chosen their
quasi-triangular structure (the universal R-matrix) gives rise to
trigonometric solutions of the Yang-Baxter equation, which have a direct
physical interpretation either as the two-particle amplitude in factorizable
scattering matrix theory or as the Boltzmann weights of a two-dimensional
statistical lattice model. In this work we will concentrate on the latter
application.

Despite this intimate relation between solvable lattice models and quantum
algebras it is important to keep in mind that in general $U_{q}(\hat{g})$
does not provide a symmetry of the physical system. In fact, the
integrability of the model follows from the Yang-Baxter equation alone and
manifests itself in commuting transfer matrices, which form an abelian
symmetry. In contrast, the quantum algebra is non-abelian and its generators
neither commute with the transfer matrix nor with the charges of the model
when periodic boundary conditions are imposed on the lattice. The latter,
however, are often chosen to render the model translation invariant
simplifying its physical discussion via the Bethe ansatz \cite{Bethe}. The
connection between the Bethe ansatz and the representation theory of the
quantum algebra $U_{q}(\hat{g})$ in the presence of periodic boundary
conditions is up-to-day not fully understood.

The situation becomes different when the deformation parameter $q$
approaches a root of unity, $q^{N}\rightarrow 1$. It was Baxter \cite{Baxter}
who first noted in the context of the XYZ model and one of its
specializations, the XXZ or six-vertex model related to $U_{q}(\widehat{sl}%
_{2})$, that in this case extra degeneracies in the eigenvalue spectrum of
the transfer matrix appear. Subsequently, the properties of the XXZ model at
roots of unity have also been investigated by several other authors \cite
{K,PS,AGR}.

The symmetry underlying these degeneracies remained unclear until Deguchi,
Fabricius and McCoy \cite{DFM,FM} showed that it can be linked to a
finite-dimensional representation of the non-deformed affine algebra $%
\widehat{sl}_{2}$ in certain commensurate sectors where the spin is a
multiple of the order $N$ of the root of unity. In particular, the symmetry
generators can be explicitly constructed as a subalgebra of $U_{q}(\widehat{%
sl}_{2})$ and are compatible with periodic boundary conditions. As recently
discussed in \cite{FM0} this allows to connect the finite-dimensional
representation theory of the affine algebra with the methods of the
algebraic Bethe ansatz leading to an exact formula for the dimension of the
degenerate eigenspaces at arbitrary roots of unity. See also \cite{BA} for a
combinatorial approach in the special case $N=6$.

The occurrence of the loop symmetry at roots of unity is a general phenomena
as has been demonstrated in \cite{KM}, where the discussion of the
six-vertex model is widened from the fundamental representation to higher
spin and generalized to arbitrary quantum affine algebras $U_{q}(\hat{g})$\
covering\ to a large extent the known integrable vertex models. In this work
we extend the discussion even further by considering integrable vertex
models whose $R$-matrix is a trigonometric solution to the graded
Yang-Baxter equation \cite{GYB}.

While from the statistical mechanics point of view no gradation is
distinguished, those algebras which carry an $\mathbb{Z}_{2}$-gradation and
are known as superalgebras have been most thoroughly discussed in the
literature, see \cite{Kac,Rit} and also \cite{FSSb} for a recent
presentation. The definition of quantum affine superalgebras and their
connection to integrable models has been developed in e.g. 
[23-32]. A posteriori it became
clear that pre-existing vertex models such as \cite{Sut,Perk} belonged to
this class. Additional motivation to consider models associated with quantum
superalgebras has been the observation that their associated spin-chain
Hamiltonians provide generalizations of the Hubbard model and describe quasi
one-dimensional strongly-correlated electron systems (see e.g. 
[35-41] for models associated with quantum
superalgebras and \cite{KE} for further references).

While we have outlined so far the physical interest in considering the root
of unity case this article will mainly be concerned with the mathematical
structure needed for the construction of the non-abelian symmetries at roots
of unity. The representation theory of quantum algebras at roots of unity
entered the mathematical literature roughly twenty years after Baxter's
observation of degeneracies in the XYZ respectively XXZ model. When $\hat{g}$
is non-graded and $q^{N}=1$ there have been two different forms of the
quantum algebra discussed, the non-restricted form $U_{q}(\hat{g})$ \cite
{CK,CKP,BK} and the restricted one $U_{q}^{\text{res}}(\hat{g})$ \cite
{Lus,CP2}. The latter is obtained when dividing the $q$-deformed
Chevalley-Serre generators $e_{i}^{n},f_{i}^{n}$ by their $q$-deformed
powers. While for generic deformation parameter $q$ the two realizations are
equivalent, they lead to quite different structures at roots of unity $%
q^{N}=1$ both of which have physical applications. The non-restricted form
is relevant for the chiral Potts model whose original formulation \cite{Au}
in the physics literature again preceded the mathematical discussion (see
also \cite{BS} for the connection with cyclic representations of $U_{q}(%
\widehat{sl}_{2})$). The physical importance of the restricted form $U_{q}^{%
\text{res}}(\hat{g})$ when $\hat{g}$ is non-graded has become apparent by
the discussion in \cite{DFM,KM} as outlined above. The symmetry generators
underlying the degeneracies at roots of unity can be directly obtained as a
subalgebra from $U_{q}^{\text{res}}(\hat{g})$.

When $\hat{g}$ is an affine superalgebra the analogous discussion of
representation theory is largely missing in the literature. While the root
of unity limit of the non-restricted quantum superalgebra $U_{q}(\hat{g})$
has been considered for some specific cases (see e.g. 
[27,50-54]) the structure of the restricted algebra $%
U_{q}^{\text{res}}(\hat{g})$ has not been subject to investigations so far.
In this article we will put forward its definition and discuss its algebraic
structure. This will put us in the position to construct explicitly
non-abelian symmetry algebras of integrable models. The detailed outline of
this paper is as follows.

In Section 2 we shortly review the definition of superalgebras and their $q$%
-deformed counterparts in order to keep the article self-contained. We will
focus on those aspects which are relevant in hindsight of our discussion and
its connection to integrable lattice models. In particular, we recall the
non-commutative Hopf algebra structure which provides the natural setting
for the graded Yang-Baxter equation.

In Section 3 we introduce analogously to the non-graded case the restricted
quantum algebra $U_{q}^{\text{res}}(\hat{g})$ for affine superalgebras $\hat{%
g}$. We calculate the supercommutation relation of its elements, state their
coproduct formulas and derive the analogue of Lustzig's higher order quantum
Serre relations \cite{Lus} for superalgebras.

In Section 4 we discuss in detail the root of unity limit and demonstrate
that the analogue of the non-graded quantum Frobenius homomorphism $U_{q}^{%
\text{res}}(\hat{g})\rightarrow U(\hat{g})$ at roots of unity holds at most
for the even subalgebras in the super case.

In Section 5 we relate the previous discussion to integrable lattice models
associated with quantum affine superalgebras. We in particular investigate
under which conditions the generators of $U_{q}^{\text{res}}(\hat{g})$ in
the $L$-fold tensor product are translation invariant when $q^{N}\rightarrow
1$. These findings then allow us to identify the subalgebras which form
non-abelian symmetry algebras of integrable lattice models.

Section 6 contains our conclusions.

\section{Quantum superalgebras}

This section gives a short review on the classification of superalgebras. We
will only consider contragredient or basic superalgebras, which possess a
non-degenerate invariant bilinear form and are the most relevant ones for
the application we have in mind. For details we refer the reader to \cite
{Kac} (see also \cite{FSSb} for a recent presentation and further references
on superalgebras). We start by recalling that superalgebras $g$ are
generalizations of Lie algebras which carry an $\mathbb{Z}_{2}$-grading
expressed in the vector space decomposition 
\begin{equation*}
g=g_{0}\oplus g_{1}\;.
\end{equation*}
Assuming that there exists an homogeneous basis one assigns to elements $x$
in the even subspace $g_{0}$ the degree $|x|=0$ while elements $y$ in the
odd subspace $g_{1}$ carry the degree $|y|=1$. The superalgebraic structure
is invoked by introducing the superbracket $[\cdot ,\cdot ]:g\times
g\rightarrow g$ obeying super antisymmetry and the super Jacobi identity for
the homogeneous elements $x,y,z\in g_{0}$ or $g_{1}$, 
\begin{eqnarray*}
\left[ x,y\right] +(-1)^{|x||y|}[y,x] &=&0 \\
(-)^{|x||z|}[[x,y],z]+(-)^{|y||x|}[\left[ y,z\right] ,x]+(-)^{|z||y|}[\left[
z,x\right] ,y] &=&0\;.
\end{eqnarray*}
Similar to the non-graded case one can classify all contragredient
superalgebras (CSA) by Cartan matrices respectively root systems $\Phi $.
One might in particular introduce the gradation of the algebra by first
grading its root system $\Phi =\Phi _{0}\cup \Phi _{1}$. Let $\left\langle
\cdot ,\cdot \right\rangle $ denote the super Killing form, then one
distinguishes the following types of roots,

\begin{itemize}
\item  if $\left\langle \alpha |\alpha \right\rangle \neq 0$ and $2\alpha
\notin \Phi $, then $\alpha \in \Phi _{0}$ is called even or white.

\item  if $\left\langle \alpha |\alpha \right\rangle \neq 0$ and $2\alpha
\in \Phi $, then $\alpha \in \Phi _{1}$ is odd and called black.

\item  if $\left\langle \alpha |\alpha \right\rangle =0$, then $\alpha \in
\Phi _{1}$ is odd and called grey.
\end{itemize}

\noindent This grading of the root system determines the grading of the
algebra by assigning to the step operators $\bar{e}_{\alpha },\bar{e}%
_{-\alpha }$ the same degree as their associated root. Note in particular
that one has $\bar{e}_{\alpha }^{2}=\bar{e}_{-\alpha }^{2}=0$ for grey
roots. The elements of the Cartan subalgebra are chosen to be even. Instead
of working with the whole root system it is convenient to choose a set of
simple roots $\Pi :=\{\alpha _{1},...,\alpha _{r}\}\subset \Phi $ and to
introduce the symmetric Cartan matrix $A_{ij}=\left\langle \alpha
_{i}|\alpha _{j}\right\rangle $ as well as the corresponding Chevalley-Serre
basis which determine the superalgebra completely.

\begin{definition}
Given a root system $\Phi =\Phi _{0}\cup \Phi _{1}$ and a set of simple
roots $\Pi \subset \Phi $ with symmetric Cartan matrix $A\in \mathbb{Z}%
^{r}\otimes \mathbb{Z}^{r}$ we assign to it the following unique
superalgebra $g=g(A,\Pi )$ whose Chevalley-Serre generators $\{h_{i},\bar{e}%
_{i}\equiv \bar{e}_{\alpha _{i}},\bar{f}_{i}\equiv \bar{e}_{-\alpha
_{i}}\}_{\alpha _{i}\in \Pi }$ obey the relations,

\begin{description}
\item[\emph{(CSA1)}]  
\begin{equation*}
\lbrack h_{i},h_{j}]=0\;,\quad \lbrack h_{i},\bar{e}_{j}]=A_{ij}\bar{e}%
_{j}\;,\quad \lbrack h_{i},\bar{f}_{j}]=-A_{ij}\bar{f}_{j}\;,\quad \lbrack 
\bar{e}_{i},\bar{f}_{j}]=\delta _{ij}h_{i}
\end{equation*}

\item[\emph{(CSA2)}]  \emph{[Chevalley-Serre relations]} 
\begin{equation*}
(\limfunc{ad}\bar{e}_{i})^{1-a_{ij}}\bar{e}_{j}=(\limfunc{ad}\bar{f}%
_{i})^{1-a_{ij}}\bar{f}_{j}=0
\quad \text{with\quad }a_{ij}:=\left\{ 
\begin{array}{cc}
2A_{ij}/A_{ii}\,, & A_{ii}\neq 0 \\ 
-1\,, & A_{ii}=0,A_{ij}\neq 0 \\ 
0\,, & \text{else}
\end{array}
\right.
\end{equation*}

\item[\emph{(CSA3)}]  In addition one has to impose extra Serre relations
whenever grey roots are present, i.e. $A_{ii}=\left\langle \alpha
_{i}|\alpha _{i}\right\rangle =0$ for some $\alpha _{i}\in \Pi $. For
example provided that 
\begin{equation*}
A_{ij}=-A_{ik}\neq 0\quad \text{and}\quad A_{ii}=A_{jk}=0
\end{equation*}
one has the relation 
\begin{equation*}
\lbrack \lbrack \bar{e}_{i},\bar{e}_{j}],[\bar{e}_{i},\bar{e}_{k}]]=[[\bar{f}%
_{i},\bar{f}_{j}],[\bar{f}_{i},\bar{f}_{k}]]=0\;.
\end{equation*}
For a complete list of the extra Serre relations, for which no universal
formula is known, we refer the reader to \cite{Yam}.
\end{description}
\end{definition}

We emphasize that in the presence of grey roots one encounters two new
features which are characteristic to superalgebras and have no analogue for
ordinary simple algebras. One of them is the occurrence of the extra Serre
relations mentioned in (CSA3). The other one concerns the existence of
inequivalent root systems for the same CSA.

If the simple root system of $g(A,\Pi )$ contains grey roots one might
generate a different simple root system $\Pi ^{\prime }$ by applying
generalized Weyl reflections associated with grey roots to $\Pi $ \cite
{serga,LSS}, 
\begin{equation}
\sigma _{i}\alpha _{j}=\alpha _{j}-a_{ij}\alpha _{i}
\end{equation}
The inequivalent root system might have a different number of odd roots and
its associated Cartan matrix $A^{\prime }$ can not be related to $A$ by a
similarity transformation in general. However, the algebras $g(A,\Pi )$ and $%
g(A^{\prime },\Pi ^{\prime })$ generated by the two different
Chevalley-Serre bases are isomorphic. There always exists a unique simple
root system, called distinguished, where the number of even simple roots is
maximal. From this root system one can construct successively all
inequivalent simple root systems, see \cite{serga} for details.

\begin{remark}
Henceforth we shall always work in the distinguished root system and its
associated Chevalley-Serre basis.
\end{remark}

We now turn to the definition of quantum superalgebras which are constructed
as $q$-deformation of the universal enveloping algebra $U(g)$. The latter is
obtained from the graded tensor algebra $\bigoplus_{n}g^{\otimes n}$ with
tensor product 
\begin{equation}
(1\otimes x)\otimes (y\otimes 1)=(-1)^{|x||y|}y\otimes x\;  \label{gt}
\end{equation}
when dividing out the ideal generated from elements of the form 
\begin{equation}
x\otimes y-(-1)^{|x||y|}y\otimes x-[x,y]\;.
\end{equation}
In other words we might identify the superbracket with the supercommutator
in $U(g)$. For convenience we drop from now on the tensor product sign $%
\otimes $ in $U(g)$. We are now prepared to define the $q$-deformation of
the universal enveloping algebra.

\begin{definition}
The \textbf{quantum universal enveloping superalgebra} $U_{q}(g)$ is the
algebra of power series in the Chevalley-Serre generators $%
\{e_{i},f_{i},h_{i}\}\cup \{1\}$ subject to the following supercommutation
relations:

\begin{description}
\item[\emph{(QSA1)}]  Let $A$ denote the symmetric Cartan matrix associated
with the superalgebra $g$. Then 
\begin{equation}
\lbrack h_{i},h_{j}]=0\quad \text{or equivalently\quad }%
q^{h_{i}}q^{-h_{i}}=q^{-h_{i}}q^{h_{i}}=1
\end{equation}
and 
\begin{equation}
q^{h_{i}}e_{j}q^{-h_{i}}=q^{A_{ji}}e_{j}\;,\quad
q^{h_{i}}f_{j}q^{-h_{i}}=q^{-A_{ji}}f_{i}\;,\quad \lbrack
e_{i},f_{j}]=\delta _{ij}\,\dfrac{q^{h_{i}}-q^{-h_{i}}}{q-q^{-1}}\;.
\end{equation}

\item[\emph{(QSA2)}]  In addition, the generators obey the quantum Serre
relations 
\begin{equation}
\left( \limfunc{ad}_{q^{\pm 1}}e_{i}\right) ^{1-a_{ij}}e_{j}=0\quad 
\text{and\quad }\left( \limfunc{ad}_{q^{\pm 1}}f_{i}\right)
^{1-a_{ij}}f_{j}=0\quad (i\neq j,\;A_{ii}\neq 0)
\end{equation}
where the $q$-deformed adjoint action is defined in terms of the $q$%
-deformed supercommutator 
\begin{equation}
\lbrack e_{\alpha },e_{\beta }]_{q^{\pm 1}}:=e_{\alpha }e_{\beta
}-(-)^{|\alpha ||\beta |}q^{\pm \left\langle \alpha |\beta \right\rangle
}e_{\beta }e_{\alpha }\;.
\end{equation}

\item[\emph{(QSA3)}]  In case that grey roots are present $(A_{ii}=0)$ there
are extra quantum Serre relations. For example, under the same conditions as
stated in (CSA3) above one has 
\begin{equation*}
\lbrack \lbrack e_{i},e_{j}]_{q^{\pm 1}},[e_{i},e_{k}]_{q^{\pm 1}}]_{q^{\pm
1}}=0\;\text{.}
\end{equation*}
Similar relations hold for the generators $f_{i}$.
\end{description}
\end{definition}

The so-defined quantum superalgebras can be endowed with the structure of a
Hopf algebra. We choose the following conventions for coproduct and
antipode, 
\begin{eqnarray}
\Delta (q^{h_{i}}) &=&q^{h_{i}}\otimes q^{h_{i}}\;,\quad \gamma
(q^{h_{i}})=q^{-h_{i}}  \notag \\
\Delta (e_{i}) &=&e_{i}\otimes q^{-\frac{h_{i}}{2}}+q^{\frac{h_{i}}{2}%
}\otimes e_{i}\;,\text{\quad }\gamma (e_{i})=-q^{-\rho }e_{i}q^{\rho } 
\notag \\
\Delta (f_{i}) &=&f_{i}\otimes q^{-\frac{h_{i}}{2}}+q^{\frac{h_{i}}{2}%
}\otimes f_{i}\;,\quad \gamma (f_{i})=-q^{-\rho }f_{i}q^{\rho }  \label{cop}
\end{eqnarray}
Here $\rho $ is the unique element in the Cartan subalgebra which satisfies $%
\alpha _{i}(\rho )=\left\langle \alpha _{i}|\alpha _{i}\right\rangle /2$ for
any simple root $\alpha _{i}$. Note that this defines in fact a graded Hopf
algebra, i.e. the coproduct preserves the grading $U_{q}(g)=U_{q}(g)_{0}%
\oplus U_{q}(g)_{1}$ inherited from $g=g_{0}\oplus g_{1}$, 
\begin{equation}
U_{q}(g)_{0}\overset{\Delta }{\rightarrow }U_{q}(g)_{0}\otimes
U_{q}(g)_{0}+U_{q}(g)_{1}\otimes U_{q}(g)_{1}
\end{equation}
and 
\begin{equation}
U_{q}(g)_{1}\overset{\Delta }{\rightarrow }U_{q}(g)_{0}\otimes
U_{q}(g)_{1}+U_{q}(g)_{1}\otimes U_{q}(g)_{0}\;.
\end{equation}
Moreover, the antipode obeys 
\begin{equation}
\gamma (xy)=(-1)^{|x||y|}\gamma (y)\gamma (x)\;.
\end{equation}
As we infer from the definition of the coproduct the quantum superalgebra $%
U_{q}(g)$ is in general non-cocommutative. In formulas this means that the
action of the \textbf{'}opposite\textbf{' }coproduct 
\begin{equation}
\Delta ^{\text{op}}\equiv \pi \circ \Delta ,  \label{opcop}
\end{equation}
does not coincide with the action of $\Delta $. Here $\pi $ denotes the
graded permutation operator, 
\begin{equation}
\pi (x\otimes y)=(-1)^{|x||y|}y\otimes x\;.  \label{pi}
\end{equation}
However, both coproduct structures can be related by an invertible element,
the universal $R$-matrix $\mathcal{R}\in U_{q}(g)\otimes U_{q}(g)$, 
\begin{equation}
\Delta ^{\text{op}}(x)=\mathcal{R}\,\Delta (x)\,\mathcal{R}^{-1}\;.
\label{R}
\end{equation}
In addition the R-matrix has to satisfy the following well known identities, 
\begin{eqnarray}
(1\otimes \Delta )\mathcal{R} &=&\mathcal{R}_{13}\mathcal{R}_{12}  \notag \\
(\Delta \otimes 1)\mathcal{R} &=&\mathcal{R}_{13}\mathcal{R}_{23}  \notag \\
(\gamma \otimes 1)\mathcal{R} &=&(1\otimes \gamma ^{-1})\mathcal{R}=\mathcal{%
R}^{-1}  \label{R1}
\end{eqnarray}
From the first two relations and the defining property of the $R$-matrix one
infers that it provides a constant solution to the graded Yang-Baxter
equation in $U_{q}(g)\otimes U_{q}(g)\otimes U_{q}(g)$, 
\begin{equation}
\mathcal{R}_{12}\mathcal{R}_{13}\mathcal{R}_{23}=\mathcal{R}_{23}\mathcal{R}%
_{13}\mathcal{R}_{12}\;.
\end{equation}
Here the lower indices indicate on which pair of copies the $R$-matrix acts.
Note that the grading is hidden in the definition of the tensor product (\ref
{gt}). In order to obtain spectral parameter dependent solutions one has to
consider affine superalgebras $\hat{g}$. The affine extensions can be
analogously defined as in the non-graded case and amount to adding one more
triplet of generators $\{e_{0},f_{0},h_{0}\}$ associated with the affine
root $\alpha _{0}$ to the Chevalley-Serre basis. For a listing of the
possible affine extensions of CSAs in terms of (affine extended) Cartan
matrices respectively Dynkin diagrams we refer the reader to e.g. \cite
{Kac,FSS,Yam}. It is then convenient to introduce the normalized or
truncated $R$-matrix 
\begin{equation}
R=q^{-c\otimes d-d\otimes c}\mathcal{R}  \label{tR}
\end{equation}
with $c$ being the central element of the affine superalgebra $\hat{g}$ and $%
d$ being the homogeneous degree operator defined by the commutation
relations 
\begin{equation}
\lbrack d,e_{i}]=\delta _{i0}e_{i}\;,\quad \lbrack d,f_{i}]=-\delta
_{i0}f_{i}\;,\quad \lbrack d,h_{i}]=0  \label{hd}
\end{equation}
and the Hopf algebra relations 
\begin{equation}
\Delta (d)=d\otimes 1+1\otimes d\;,\quad \gamma (d)=-d\;.
\end{equation}
The truncated $R$-matrix can now be given a spectral parameter dependence by
introducing the automorphism 
\begin{equation}
D_{z}(x)=z^{d}\,x\,z^{-d}\;,\quad z\in \mathbb{C},\;x\in U_{q}(\hat{g})\;.
\end{equation}
and setting 
\begin{equation}
R(z)=(D_{z}\otimes 1)R=(1\otimes D_{z^{-1}})R\;.  \label{zR}
\end{equation}
From the above equation for the universal R-matrix we now infer that the
spectral parameter dependent R-matrix satisfies the equation 
\begin{equation}
R_{12}(z)R_{13}(zwq^{1\otimes c\otimes
1})R_{23}(w)=R_{23}(w)R_{13}(zwq^{-1\otimes c\otimes 1})R_{12}(z)
\label{gyb}
\end{equation}
which is the known form of statistical mechanics respectively factorizable
S-matrix theory provided we choose a representation where the central
element $c$ is set to zero. The other relations enjoyed by the universal
R-matrix translate to 
\begin{eqnarray}
(\Delta _{z}\otimes 1)R(w) &=&R_{13}(zwq^{1\otimes c\otimes 1})R_{23}(w) 
\notag \\
(1\otimes \Delta _{z})R(w) &=&R_{13}(wz^{-1}q^{-1\otimes c\otimes
1})R_{12}(w)  \notag \\
(\gamma \otimes 1)R(z) &=&R(zq^{c\otimes 1})^{-1}  \notag \\
(1\otimes \gamma ^{-1})R(z) &=&R(zq^{-1\otimes c})^{-1}
\end{eqnarray}
as well as the quasi-triangular property 
\begin{equation}
R(z)\Delta _{z}(x)=q^{-d\otimes c-c\otimes d}\Delta _{z}^{\text{op}%
}(x)q^{d\otimes c+c\otimes d}\,R(z)  \label{jime}
\end{equation}
which amounts to Jimbo's celebrated equations \cite{Jimbo} for the
construction of R-matrices. Again we emphasize that we are going to consider
only finite-dimensional representations $\rho _{V}:U_{q}(\hat{g})\rightarrow 
\mathrm{End}(V)$ where $c=0$. Then the above $q$-factors drop out.

\section{The restricted algebra}

In this section we discuss the restricted quantum algebra $U_{q}^{\text{res}%
}(\hat{g})$ which will be similarly defined as in the non-graded case
(compare \cite{Lus}). We recall from the non-graded case that the
representation theory of $U_{q}^{\text{res}}(\hat{g})$ as opposed to the
non-restricted form $U_{q}(\hat{g})$ in the root of unity limit is different
(see \cite{Lus,CP2}\ and \cite{BK}). As shown in \cite{DFM,KM} for
non-graded affine algebras the relevant structure for integrable models is
given by $U_{q}^{\text{res}}(\hat{g})$ whence we define here the analogue of
this particular realization for quantum affine superalgebras.

\begin{definition}
The restricted quantum superalgebra$\,U_{q}^{\text{res}}(\hat{g})$ is the
algebra generated by the elements 
\begin{equation}
\{e_{i}^{(n)},f_{i}^{(n)},q^{h_{i}},q^{-h_{i}}\}_{n\in \mathbb{N}}\;\text{%
with}\quad e_{i}^{(n)}:=\frac{e_{i}^{n}}{[n]_{q_{i}}!}\,,\;f_{i}^{(n)}:=%
\frac{f_{i}^{n}}{[n]_{q_{i}}!}\;.  \label{res}
\end{equation}
Here we have introduced the following super $q$-integers associated with a
simple root $\alpha _{i}$, 
\begin{equation}
\lbrack n]_{q_{i}}:=\frac{(-)^{|i|n}q_{i}^{n}-q_{i}^{-n}}{%
(-)^{|i|}q_{i}-q_{i}^{-1}}=(-)^{|i|(n-1)}[n]_{q_{i}^{-1}}\;,\quad q_{i}:=q^{%
\frac{\left\langle \alpha _{i}|\alpha _{i}\right\rangle }{2}}  \label{qint}
\end{equation}
and the corresponding factorials as well as binomial coefficients 
\begin{equation}
\lbrack m]_{q_{i}}!:=\prod_{n=1}^{m}[n]_{q_{i}}\;,\quad \qatopd{m}{n}%
_{q_{i}}:=\frac{[m]_{q_{i}}!}{\left[ n\right] _{q_{i}}!\left[ m-n\right]
_{q_{i}}!}=(-)^{|i|n(m-n)}\qatopd{m}{n}_{q_{i}^{-1}}\;.
\end{equation}
\end{definition}

To motivate this definition we recall from the non-graded case that it is
crucial to divide the generators by the $q$-integers in order to obtain a
non-trivial root of unity limit, $q^{N}\rightarrow 1$, for the $N^{\text{th}%
} $ power of the step operators $e_{i},f_{i}$ which otherwise would become
central. Note that for even roots (\ref{res}) reduces to the non-graded
restricted quantum algebra since the super $q$-integers (\ref{qint}) turn
into ordinary $q$-integers $\left\lfloor n\right\rfloor _{q}=\left(
q^{n}-q^{-n}\right) /\left( q^{1}-q^{-1}\right) $. In fact, one might define
the super $q$-integers in terms of the ordinary ones by the following
relation, 
\begin{equation}
\lbrack n]_{q_{i}}=\left( \sqrt{-1}\right) ^{|i|(n-1)}\left\lfloor
n\right\rfloor _{\bar{q}_{i}}\quad \text{with\quad }\bar{q}_{i}=\left( \sqrt{%
-1}\right) ^{|i|}q_{i}\;.
\end{equation}
As an immediate consequence of this relation one has the following lemma.

\begin{lemma}
The super $q$-integers satisfy the following three identities, 
\begin{gather}
\lbrack n]_{q_{i}}=q_{i}^{-n+1}\sum\limits_{l=0}^{n-1}(-)^{|i|l}q_{i}^{2l}\;,
\label{1} \\
\lbrack m+n]_{q_{i}}=(-)^{n|i|}q_{i}^{n}[m]_{q_{i}}+q_{i}^{-m}[n]_{q_{i}}\;,
\label{2} \\
\prod\limits_{k=0}^{n-1}(1+(-)^{|i|k}q_{i}^{2k}z)=\sum%
\limits_{l=0}^{n}(-)^{|i|\frac{l(l-1)}{2}}q_{i}^{l(n-1)}\qatopd{n}{l}%
_{q_{i}}z^{l}\;.  \label{3}
\end{gather}
\end{lemma}

These formulas together with the definition of the super $q$-integers will
not only simplify the calculations considerably but also allow us to write
down compact and elegant formulas, since the additional introduced sign
factors in (\ref{qint}) will turn out to reflect conveniently the grading of
the superalgebra.

Having defined the restricted algebra $U_{q}^{\text{res}}(\hat{g})$ we
consider the supercommutator of its generators. One obtains by induction the
following formula, whose proof can be found in the appendix, 
\begin{equation}
\lbrack e_{i}^{(m)},f_{i}^{(n)}]=\sum_{k=1}^{\min
(m,n)}(-)^{|i|(m-k)(n-k)}f_{i}^{(n-k)}e_{i}^{(m-k)}\prod_{l=1}^{k}\frac{%
\left[ h_{i};m-n-l+1\right] }{[l]_{q_{i}}}\;.  \label{F1}
\end{equation}
Here we have introduced the Cartan elements 
\begin{equation}
\lbrack h_{i};m]:=\frac{q^{h_{i}}q_{i}^{m}-(-)^{|i|m}q^{-h_{i}}q_{i}^{-m}}{%
q-q^{-1}}\;.
\end{equation}
These formulas generalize the ones obtained in \cite{Lus,CK} to the quantum
superalgebra case and will be repeatedly used when we discuss the root of
unity limit. We also rewrite the quantum Serre relations. By induction one
proves (see appendix) that the $m$-fold $q$-deformed adjoint action can be
explicitly written out as follows, 
\begin{equation}
\left( \mathrm{ad}_{q^{\pm 1}}e_{i}\right)
^{m}e_{j}^{n}=\sum_{s=0}^{m}(-1)^{s(1+n|i||j|)+|i|\frac{s(s-1)}{2}%
}q_{i}^{\mp s(1-na_{ij}-m)}\qatopd{m}{s}_{q_{i}^{\pm
1}}\,e_{i}^{m-s}e_{j}^{n}e_{i}^{s}
\end{equation}
Setting $m=1-a_{ij},n=1$ one obtains the quantum Serre relations stated in
the definition (QSA2). Using the general formula for arbitrary $m,n$ we now
derive the analogue of Lustzig's quantum higher order Serre relations \cite
{Lus} for superalgebras.

\begin{proposition}
Define the following element in $U_{q}(\hat{g})$, 
\begin{equation}
\Theta _{m,n}:=\frac{\left( \limfunc{ad}_{q}e_{i}\right)
^{m}e_{j}^{n}}{[m]_{q_{i}}![n]_{q_{i}}!}%
=\sum_{r+s=m}(-)^{s+x_{s}^{n}}q_{i}^{-s(1-na_{ij}-m)}e_{i}^{(r)}e_{j}^{(n)}e_{i}^{(s)}
\label{def}
\end{equation}
with the degree function set to 
\begin{equation}
x_{s}^{n}:=|i|\frac{s(s-1)}{2}+|i||j|ns\;.  \label{x}
\end{equation}
Then provided that $m\geq 1-n\,a_{ij}>0$ one has 
\begin{equation}
\Theta _{m,n}=0\;.
\end{equation}
We shall refer to these identities as the higher order quantum Serre
relations.
\end{proposition}

\begin{proof}
The proof proceeds by induction along the lines for the non-graded case. For 
$n=1$ and $m=1-na_{ij}$ we obtain the ordinary quantum Serre relations as
already seen above. Assuming that the assertion is true for some $n-1,n>1$
one performs the induction step by means of the following identities which
can be verified by direct calculation.

Setting $m_{o}:=1-na_{ij}$ one has 
\begin{equation}
\lbrack \Theta _{m_{o},n},f_{i}]=\left[ 1-(-)^{|i|na_{ij}}\right] \frac{%
q^{h_{i}}}{q-q^{-1}}\Theta _{m_{o}-1,n}=0\;.
\end{equation}
Note that the vanishing of the supercommutator only follows if $|i|=0$ or $%
a_{ij}=0\func{mod}2$. The first case is trivial. Suppose that $|i|=1$ then a
quick study of the Cartan matrices or the Dynkin diagrams reveals that $%
a_{ij}$ with $\alpha _{i}$ being a black simple root is satisfied for all
Kac-Moody superalgebras.

The second identity we exploit is given by 
\begin{equation}
\lbrack \Theta _{m_{o},n},f_{j}]=(-)^{|j|(n-1)}\left\{ \tfrac{%
q^{-h_{j}}q_{j}^{-\left( 1-m_{o}a_{ji}-n\right) }}{q^{-1}-q}\Theta
_{m_{o},n-1}\right\} +(-)^{|j|\frac{n(n-1)}{2}}\left\{ \,q\rightarrow
q^{-1}\right\},
\end{equation}
where the last term in brackets is obtained from the first one when
replacing $q$ by $q^{-1}$. Also this supercommutator vanishes by the
induction hypothesis.

Thus, we conclude that $\Theta _{m_{o},n}$ supercommutes with $f_{i},f_{j}$
and therefore trivially with all $f_{k},k=0,1,...,r$. Hence, it must vanish.
This completes the induction proof for $m=m_{o}$. Using the defining
relation of $\Theta _{m,n}$ we see that the assertion also follows for
larger $m$.
\end{proof}

For later purposes it will be useful to rewrite these higher order quantum
Serre relations in a different form.

\begin{corollary}
For $m\geq 1-n\,a_{ij}>0$ $\left( |i|=0\text{ or }a_{ij}=0\func{mod}2\right) 
$ the above relations can be rewritten as 
\begin{equation}
0=e_{i}^{(m)}e_{j}^{(n)}+\sum_{\substack{ r+s=m  \\ m+na_{ij}\leq s\leq m}}%
c_{s}\,e_{i}^{(r)}e_{j}^{(n)}e_{i}^{(s)}\;  \label{HOS}
\end{equation}
with the coefficient function equal to 
\begin{equation}
c_{s}:=(-)^{s+x_{s}^{n}}q_{i}^{-s(1-na_{ij}-m)}%
\sum_{p=0}^{m+na_{ij}-1}(-)^{p+|i|\frac{p(p-1)}{2}}q_{i}^{-p(s-1)}\qatopd{s}{p}_{q_{i}}\;.  \label{cs}
\end{equation}
Here the degree function $x_{s}^{n}$ is defined as in (\ref{x}).
\end{corollary}

\begin{proof}
First one observes that for $0\leq p\leq m+na_{ij}-1$ one has 
\begin{equation*}
0=\sum_{p=0}^{m+na_{ij}-1}(-)^{|i||j|np}q_{i}^{p(na_{ij}+m-p)}\Theta
_{m-p,n}e_{i}^{(p)}\;.
\end{equation*}
Plugging in the definition of $\Theta _{m,n}$ and exploiting formula (\ref{3}%
) for $z=-1$ the assertion follows.
\end{proof}

Since we will have below to consider higher tensor products of the
restricted algebra in order to make contact to integrable lattice models we
conclude this section by stating the coproduct formulas for $U_{q}^{\text{res%
}}(\hat{g})$. Taking into account that the coproduct (\ref{cop}) is an Hopf
algebra homomorphism one obtains by induction 
\begin{equation}
\Delta (e_{i}^{(m)})=\sum_{n=0}^{m}\,e_{i}^{(n)}q^{(m-n)\frac{h_{i}}{2}%
}\otimes e_{i}^{(m-n)}q^{-n\frac{h_{i}}{2}}\;.  \label{L2}
\end{equation}
A similar formula holds for the generators $f_{i}^{(n)}$. At first sight
this formula does not seem to differ from the non-graded case. However, we
point out that the graded tensor product (\ref{gt}) has to be taken into
account and that its structure is conveniently hidden in the definition of
the super $q$-integers (\ref{qint}). In order to construct higher tensor
products one defines iteratively the $L$-fold coproduct via 
\begin{equation}
\Delta ^{(L)}=\left( \Delta \otimes 1\right) \Delta ^{(L-1)}\quad \text{%
with\quad }\Delta ^{(2)}\equiv \Delta \;.
\end{equation}
Taking formula (\ref{L2}) as a starting point and exploiting once more the
fact that the coproduct is an algebra homomorphism one arrives at 
\begin{equation}
\Delta ^{(L)}(e_{i}^{(m)})=\sum_{0=n_{0}\leq \ldots \leq
n_{L}=m}\,\bigotimes_{l=1}^{L}e_{i}^{\left( n_{l}-n_{l-1}\right) }q^{\left(
m-n_{l}-n_{l-1}\right) \frac{h_{i}}{2}}\;,
\end{equation}
Analogous formulas hold for the generators $f_{i}$. Again we stress that
this formula resembles closely the one in the non-graded case since the
grading is conveniently encoded in the definition of the super $q$-integers.

\section{The root of unity limit of $U_{q}^{res}(\hat{g})$}

In this section we discuss the root of unity limit $q^{N}\rightarrow 1$ and
henceforth we assume that $q^{N}=1$ is primitive. Since the outcome will
depend crucially on the grading of the generators we discuss first the
restricted subalgebra $U_{q}^{\text{res}}(\hat{g})_{i}$ associated with a
single simple root $\alpha _{i}$ which is either white, black or grey. $%
U_{q}^{\text{res}}(\hat{g})_{i}$ is generated by the elements 
\begin{equation}
\{e_{i}^{(n)},f_{i}^{(n)},q^{h_{i}},q^{-h_{i}}\}_{n\in \mathbb{N}}\;\text{%
with}\quad e_{i}^{(n)}:=\frac{e_{i}^{n}}{[n]_{q_{i}}!}\,,\;f_{i}^{(n)}:=%
\frac{f_{i}^{n}}{[n]_{q_{i}}!}\;.
\end{equation}
As mentioned in the previous section it is important to divide the
generators by the super $q$-integers. From the non-graded case it is known
that the elements $e_{i}^{N^{\prime }},f_{i}^{N^{\prime }},q^{\pm h_{i}}$ ($%
\alpha _{i}$ white) become central in the root of unity limit $%
q^{N}\rightarrow 1$ and in non-cyclic representations take the values $%
e_{i}^{N^{\prime }},f_{i}^{N^{\prime }}=0,q^{\pm h_{i}}=\pm 1$ \cite{BK},
where the integer $N^{\prime }$ is defined as 
\begin{equation}
N^{\prime }:=\left\{ 
\begin{array}{cc}
N\;, & N\text{ odd} \\ 
N/2\;, & N\text{ even}
\end{array}
\right. \quad .
\end{equation}
In contrast the restricted generators $e_{i}^{(N^{\prime
})},f_{i}^{(N^{\prime })}$ stay well defined due to a simultaneous vanishing
of the $q$-integer $[N^{\prime }]_{q_{i}}=0$ \cite{Lus}. We will now
consider the analogue of these non-cyclic representations for arbitrary
roots $\alpha _{i}$ and shall impose the preliminary commensurability
condition 
\begin{equation}
\lambda (h_{i})=0\;\mathrm{mod}\;N^{\prime }\;\Leftrightarrow
\;q^{h_{i}}=\pm 1  \label{commp}
\end{equation}
with $\lambda $ denoting the highest weight determining the representation.
The above commensurability condition is preliminary, since we will have to
strengthen it later on when discussing translation invariance of the
restricted algebra for higher tensor products.

We are interested in the supercommutation relations of the restricted
generators at roots of unity. From the general formula (\ref{F1}) of the
previous section we see that for $m=n=N^{\prime }$ the product in the
supercommutator always contains a zero for each summand in the root of unity
limit. Therefore, its expression simplifies to 
\begin{multline}
\lim_{q^{N}\rightarrow 1}[e_{i}^{(N^{\prime })},f_{i}^{(N^{\prime
})}]=\lim_{q^{N}\rightarrow 1}\frac{[h_{i};0]}{[N^{\prime }]_{q_{i}}}%
\prod_{l=1}^{N^{\prime }-1}\frac{[h_{i};-l]}{[l]_{q_{i}}}  \notag \\
=\left( \tfrac{(-)^{|i|}q_{i}-q_{i}^{-1}}{q-q^{-1}}\right) ^{N^{\prime
}}\prod_{l=1}^{N^{\prime }-1}\frac{%
q^{h_{i}}q_{i}^{-l}-(-)^{|i|l}q^{-h_{i}}q_{i}^{l}}{%
(-)^{|i|l}q_{i}^{l}-q_{i}^{-l}}\lim_{q^{N}\rightarrow 1}\frac{%
q^{h_{i}}-q^{-h_{i}}}{(-)^{|i|N^{\prime }}q_{i}^{N^{\prime
}}-q_{i}^{-N^{\prime }}}
\end{multline}
Note that the supercommutator is only non-zero if the super $q$-integer $%
[N^{\prime }]_{q_{i}}$ vanishes. We will now evaluate the limit in the
second line of the above equation for the different cases of $\alpha _{i}$
being a white, black or grey root.

\begin{description}
\item[\textbf{$\boldsymbol{\alpha_{i}}$ white.}]  In this case all
supercommutators and super $q$-integers reduce to ordinary commutators and $%
q $-integers respectively. The discussion of the root of unity limit follows
along the lines in \cite{KM}. Setting $\left\langle \alpha _{i}|\alpha
_{i}\right\rangle =2$ one finds that 
\begin{equation}
\lim_{q^{N}\rightarrow 1}[e_{i}^{(N^{\prime })},f_{i}^{(N^{\prime
})}]=(-)^{N^{\prime }-1}q^{N^{\prime }h_{i}}\frac{h_{i}}{N^{\prime }}\;.
\label{w1}
\end{equation}
Together with the obvious commutation relations 
\begin{equation}
\lbrack h_{i},e_{i}^{(N^{\prime })}]=2N^{\prime }e_{i}^{(N^{\prime })}\quad 
\text{and\quad }[h_{i},f_{i}^{(N^{\prime })}]=-2N^{\prime }f_{i}^{(N^{\prime
})}\;  \label{w2}
\end{equation}
we conclude that to every white root we obtain a non-deformed $U(sl_{2})$
algebra in the root of unity limit generated by $\{e_{i}^{(N^{\prime
})},f_{i}^{(N^{\prime })},h_{i}/N^{\prime }\}$. Note that for odd roots of
unity we always have the real form $sl_{2}(\mathbb{R})$, while for even
roots of unity one might also obtain $sl_{2}(\mathbb{C})$.

\item[\textbf{$\boldsymbol{\alpha_{i}}$ black.}]  For black roots we infer
from the definition of the super $q$-integers that for odd roots of unity $%
q^{N}=1 $ the $q$-integer $[N]_{q_{i}}$ does not necessarily vanish. Let us
therefore first consider the case of taking an even primitive root of unity $%
q^{N}=1$, with $N^{\prime }$ even$\;$($q=e^{i\pi \frac{2k-1}{N^{\prime }}%
},k\in \mathbb{N}$). Then obviously $q^{\frac{N^{\prime }}{2}}=\pm \sqrt{-1}$
and 
\begin{equation}
\lbrack N^{\prime }/2]_{q_{i}}=\frac{q^{\frac{N^{\prime }}{2}}+q^{-\frac{%
N^{\prime }}{2}}}{q^{1}+q^{-1}}=0\;.
\end{equation}
Considering the sector $\lambda (h_{i})=0\;\mathrm{mod}\;N^{\prime }$ one
arrives at the supercommutator 
\begin{equation}
\lim_{q^{N}\rightarrow 1}[e_{i}^{(\frac{N^{\prime }}{2})},f_{i}^{(\frac{%
N^{\prime }}{2})}]=q^{\frac{N^{^{\prime }}}{2}}\,q^{-h_{i}}\left( \frac{%
q^{1}+q^{-1}}{q^{1}-q^{-1}}\right) ^{\frac{N^{\prime }}{2}}\,\frac{2h_{i}}{%
N^{\prime }}\,\;.  \label{b1}
\end{equation}
Moreover, the supercommutation relations with $h_{i}$ analogous to the white
case hold, 
\begin{equation}
\lbrack h_{i},e_{i}^{(\frac{N^{\prime }}{2})}]=N^{\prime }\,e_{i}^{(\frac{%
N^{\prime }}{2})}\quad \text{and\quad }[h_{i},f_{i}^{(\frac{N^{\prime }}{2})}%
]=-N^{\prime }\,f_{i}^{(\frac{N^{\prime }}{2})}\;.  \label{b1a}
\end{equation}
After a suitable renormalization of the step operators these
supercommutation relations resemble closely the ones of an (non-deformed) $%
osp(2|1)$ subalgebra. However, the factor $q^{-h_{i}}$ in (\ref{b1})
produces in general alternating signs in a multiplet as is immediate to see
by means of the above commutation relations with the step operators.

Therefore, we consider instead the even generators $e_{i}^{(N^{\prime
})},f_{i}^{(N^{\prime })}$ which obey 
\begin{eqnarray}
\lim_{q^{N}\rightarrow 1}[e_{i}^{(N^{\prime })},f_{i}^{(N^{\prime })}]
&=&-\left( \frac{q^{1}+q^{-1}}{q^{1}-q^{-1}}\right) ^{N^{\prime
}}q^{(N^{\prime }-1)h_{i}}\lim\limits_{q^{N}\rightarrow 1}\frac{%
q^{h_{i}}-q^{-h_{i}}}{q^{N^{\prime }}-q^{-N^{\prime }}}  \notag \\
&=&-\left( \frac{q^{1}+q^{-1}}{q^{1}-q^{-1}}\right) ^{N^{\prime
}}\,q^{N^{\prime }h_{i}}\frac{h_{i}}{N^{\prime }}  \label{b2}
\end{eqnarray}
and 
\begin{equation}
\lbrack h_{i},e_{i}^{(N^{\prime })}]=2N^{\prime }\,e_{i}^{(N^{\prime
})}\quad \text{and\quad }[h_{i},f_{i}^{(N^{\prime })}]=-2N^{\prime
}\,f_{i}^{(N^{\prime })}\;.  \label{b2a}
\end{equation}
From the last commutation relations one deduces that the factor in front of $%
h_{i}/N^{\prime }$ in equation (\ref{b2}) always stays positive. Hence, the
bosonic generators associated with a black root yield after a suitable
renormalization again an $sl_{2}(\mathbb{R})$ algebra.

Analogously, we might consider for roots of unity $q^{N}=1$ with $N^{\prime
} $ even or odd the even generators $e^{(2N^{\prime })},f^{(2N^{\prime })}$
which satisfy the commutation relations 
\begin{eqnarray}
\lim_{q^{N}\rightarrow 1}[e_{i}^{(2N^{\prime })},f_{i}^{(2N^{\prime })}]
&=&-q^{h_{i}}\left( \frac{q^{1}+q^{-1}}{q^{1}-q^{-1}}\right) ^{2N^{\prime
}}\lim\limits_{q^{N}\rightarrow 1}\frac{q^{h_{i}}-q^{-h_{i}}}{q^{2N^{\prime
}}-q^{-2N^{\prime }}}  \notag \\
&=&-\left( \frac{q^{1}+q^{-1}}{q^{1}-q^{-1}}\right) ^{2N^{\prime }}\,\frac{%
h_{i}}{2N^{\prime }}\;.  \label{b3}
\end{eqnarray}
and 
\begin{equation}
\lbrack h_{i},e_{i}^{(2N^{\prime })}]=4N^{\prime }\,e_{i}^{(2N^{\prime
})}\quad \text{and\quad }[h_{i},f_{i}^{(2N^{\prime })}]=-4N^{\prime
}\,f_{i}^{(2N^{\prime })}\;.  \label{b3a}
\end{equation}
After a suitable renormalization of the step operators we obtain once more
an $sl_{2}(\mathbb{R})$ subalgebra.

\item[\textbf{$\boldsymbol{\alpha_{i}}$ grey.}]  In this case the generators
are idempotent, $e_{i}^{2},f_{i}^{2}=0$ and the above considerations cannot
be applied.
\end{description}

\subsection{The quantum Frobenius homomorphism}

Having analyzed the subalgebras associated with a single simple root $\alpha
_{i}$ in the limit $q^{N}\rightarrow 1$, we now address the remaining
algebraic structures, namely the supercommutation and Serre relations
involving generators of different roots. We recall from the non-graded case
that the relations of the restricted algebra $U_{q}^{\text{res}}(\hat{g})$
at roots of unity can be partially identified with the relations of the
non-deformed algebra $U(\hat{g})$ via Lustzig's quantum Frobenius
homomorphism.

From the results of our case-by-case discussion of $U_{q}^{\text{res}}(\hat{g%
})_{i}$ with $\alpha _{i}$ being either white, black or grey we anticipate
that for affine superalgebras we will have restrict ourselves at least to
the even (or bosonic) subalgebra in order to give a well defined analogue of
the quantum Frobenius mapping for the graded case \cite{Lus}.

\begin{theorem}
Let $\hat{g}$ be a affine superalgebra and denote by $\hat{g}^{\text{trunc}}$
the algebra which is obtained by deleting from the Dynkin diagram of $\hat{g}
$ the nodes (and their adjoint edges) corresponding to grey roots. Then at
roots of unity $q^{N}=1$ with $N$ odd the mapping $F:U_{q}^{\text{res}}(\hat{%
g})\rightarrow U(\hat{g}_{0}^{\text{trunc}})$ defined by $F(q^{h_{i}})=1 $
and 
\begin{equation}
F(e_{\pm \alpha _{i}}^{(m)})=\left\{ 
\begin{array}{cc}
\bar{e}_{\pm \alpha _{i}}^{m/N}/(m/N)!\,,\; & m=0\func{mod}N,\alpha _{i}%
\text{ white} \\ 
\bar{e}_{\pm \alpha _{i}}^{m/2N}/(m/2N)!\,,\; & m=0\func{mod}2N,\alpha _{i}%
\text{ black} \\ 
0\,,\; & \text{else}
\end{array}
\right.\quad (e_{-\alpha_i}\equiv f_i)
\end{equation}
is an Hopf algebra homomorphism. Here $\hat{g}_{0}^{\text{trunc}}$ denotes
the even subalgebra of $\hat{g}^{\text{trunc}}$. For examples see Table 4.1.
Note in particular that in case that there are no grey roots one has an
homomorphism $U_{q}^{\text{res}}(\hat{g})\rightarrow U(\hat{g}_{0})$ with $%
\hat{g}_{0}$ being the even subalgebra of $\hat{g}$.
\end{theorem}

\begin{center}
\begin{tabular}{|l|l|l|}
\hline\hline
$\hat{g}$ & $\hat{g}^{\text{trunc}}$ & $\hat{g}_{0}^{\text{trunc}}$ \\ 
\hline\hline
$osp(2m|2n)$ & $sl_{n}\oplus so_{2m+1}$ & $sl_{n}\oplus so_{2m+1}$ \\ 
\hline\hline
$sl(2n|2n)$ & $sl_{n}\oplus sl_{n}$ & $sl_{n}\oplus sl_{n}$ \\ \hline\hline
$G(3)$ & $G_{2}$ & $G_{2}$ \\ \hline\hline
$osp(2|2n)$ & $osp(2|2n)$ & $sp_{2n}$ \\ \hline\hline
$osp(2m|2n)^{(1)}$ & $sp_{2n}\oplus so_{2m+1}$ & $sp_{2n}\oplus so_{2m+1}$
\\ \hline\hline
$osp(2|2)^{(2)}$ & $osp(2|2)^{(2)}$ & $sl_{2}^{(1)}$ \\ \hline\hline
$osp(2|2n)^{(1)}$ & $osp(2|2n)^{(1)}$ & $sp_{2n}^{(1)}$ \\ \hline\hline
\end{tabular}
\smallskip
\end{center}

\noindent {\small Table 4.1. Listed are various examples of Kac-Moody
superalgebras and the associated truncated algebra obtained by deleting the
grey nodes in the Dynkin diagram.}\medskip

\begin{proof}
For the subalgebras generated by the Chevalley-Serre basis
associated with white simple roots we can apply the results of the
non-graded case \cite{KM}. Recall that in general the correct
Chevalley-Serre relations could only be obtained for odd roots of unity in
this case, whence we restrict ourselves to $N$ odd. When dealing with the
bosonic generators $e_{i}^{(2N)},f_{i}^{(2N)}$ obtained from a black root $%
\alpha _{i}$ we first of all immediately verify that 
\begin{eqnarray*}
\lbrack h_{i},e_{j}^{(N)}] &=&NA_{ij}e_{j}^{(N^{\prime })}\quad \text{%
and\quad }[h_{j},e_{i}^{(2N)}]=2NA_{ij}e_{j}^{(N^{\prime })}\quad (\alpha
_{j}\text{ white}) \\
\lbrack h_{i},e_{j}^{(2N)}] &=&2NA_{ij}e_{j}^{(2N)}\quad \text{and\quad }[%
h_{j},e_{i}^{(2N)}]=2NA_{ij}e_{j}^{(N)}\quad (\alpha _{j}\text{ black})\;.
\end{eqnarray*}
The second case only occurs for $\hat{g}=osp(2|2)^{(2)}$. It remains to show
that the Serre relations in $U(\hat{g}_{0})$ are satisfied. For this purpose
we will exploit the higher quantum Serre relations for superalgebras proven
in the previous section.

Starting from the variant (\ref{HOS}) one infers that only the cases with $%
\alpha _{i}$ black need to be considered. For $\alpha _{i}$ white the proof
follows along the lines in the non-graded case \cite{KM}. Let us start with $%
|i|=1,|j|=0$ and set $m=2N(1-\frac{a_{ij}}{2})$ and $n=N$. Remember that for
black roots $\alpha _{i}$ one always has $a_{ij}=0\func{mod}2$. Exploiting
the algebraic properties of the super $q$-integers and taking into account
that $q^{N}=1$ one derives the following limit of the coefficient function (%
\ref{cs}), 
\begin{equation*}
\lim_{q^{N}\rightarrow 1}c_{s}=\left\{ 
\begin{array}{ll}
(-)^{\frac{s}{2N}}, & s=0\func{mod}2N \\ 
0, & \text{else}
\end{array}
\right. \;.
\end{equation*}
Taking this result together with the identity 
\begin{equation*}
\lim_{q^{N}\rightarrow 1}\frac{[2N]_{q_{i}}!^{s}}{[2Ns]_{q_{i}}!}=\frac{1}{s!%
}\;
\end{equation*}
we obtain from the higher order quantum Serre relations of $U_{q}^{\text{res}%
}(\hat{g})$ the non-deformed Serre relations 
\begin{equation*}
0=\sum_{r+s=1-\frac{a_{ij}}{2}}(-)^{s}\binom{1-\frac{a_{ij}}{2}}{s}%
e_{i}^{(2N)r}e_{j}^{(N)}e_{i}^{(2N)s}\;.
\end{equation*}
Again we remind the reader that $a_{ij}/2$ is an integer for all
superalgebras provided $i$ labels a black and $j$ a black or white root. For
example, if $\hat{g}=osp(2|2n)$ one has $a_{nn-1}=2$ and the above identity
with $a_{nn-1}\rightarrow a_{nn-1}/2$ derived in the root of unity limit
then just corresponds to the Serre relations of $U(\hat{g}_{0}=sp_{2n})$.
Note that we have defined the Chevalley-Serre basis in terms of the
symmetric Cartan matrix.

Similar one obtains for the remaining case that $|i|=|j|=1$ ($\hat{g}%
=osp(2|2)^{(2)}$) by setting $m=2N(1-a_{ij}),n=2N$ from (\ref{HOS}) the
relations 
\begin{equation*}
0=\sum_{r+s=1-a_{ij}}(-)^{s}\binom{1-a_{ij}}{s}%
e_{i}^{(2N)r}e_{j}^{(2N)}e_{i}^{(2N)s}\;.
\end{equation*}
As one easily verifies these are also the correct Serre relations of $\hat{g}%
_{0}=sl_{2}^{(1)}$. We point out that in this special case the theorem might
be extended to even roots of unity with $N^{\prime }$ odd as well, since $%
osp(2|2)^{(2)}$ contains only black simple roots.

That the homomorphism is also compatible with the coproduct structures can
be easily seen from the formula (\ref{L2}). This completes the proof.
\end{proof}

\begin{remark}
Note that for generic $q$ the even subalgebra is modified under the $q$%
-deformation. For example, while for the non-deformed algebra $U(\hat{g})$ ($%
q=1$) the squared Chevalley-Serre generators $\bar{e}_{i}^{2},\bar{f}%
_{i}^{2} $ with $\alpha _{i}$ black give rise to an $sl_{2}$ subalgebra, it
is not true that the $q$-deformed generators $e_{i}^{2},f_{i}^{2}$ of $U_{q}(%
\hat{g})$ obey the $U_{q}(sl_{2})$ commutation relations. As we have shown
in the previous section and in the above theorem one recovers the some of
the ''classical'' relations in the root of unity limit from the restricted
quantum superalgebra $U_{q}^{\text{res}}(\hat{g})$.
\end{remark}

One might wonder if the restriction to consider only the truncated affine
superalgebra $\hat{g}^{\text{trunc}}$ might be lifted and the above theorem
can be extended to the whole superalgebra as it is the case when grey roots
are absent. In order to settle this issue one would need to have an analogue
of the Cartan-Weyl basis for the quantum case. This would allow to
investigate whether there are additional bosonic generators obtained from
multiple ($q$-deformed) supercommutators involving grey step operators.
While such quantum Cartan-Weyl basis has been proposed in the literature 
\cite{QSA3,QSA5} it is not clear that it will give rise to the correct root
space structure in the root of unity limit. Moreover, since the Cartan-Weyl
generators are defined in terms of multiple ($q$-deformed) supercommutators
their coproduct structure is quite intricate. For our physical application
in the subsequent section, however, we have to consider higher tensor
products. Since we are primarily interested in the physical consequences we
leave the issue of a possible extension of the above theorem to include also
grey roots to future work.

\section{Translation invariance and symmetries of integrable lattice models}

In this section we turn to the physical application of our previous
discussion. We shortly review how to each quantum affine superalgebra a
lattice model can be assigned and then demonstrate which of the restricted
subalgebras treated before are translation invariant and form a symmetry
algebra.

Suppose we are given a finite dimensional representation $\rho _{V}:U_{q}(%
\hat{g})\rightarrow \mathrm{End}(V)$ of the quantum affine superalgebra $%
U_{q}(\hat{g})$ over some graded vector space $V=V^{(0)}\oplus V^{(1)}$.
Taking an $L\times M$ square lattice we assign to each link of the lattice a
copy of the representation space $V$ and to each vertex the spectral
parameter dependent R-matrix evaluated in this representation, $%
R^{VV}(z)=(\rho _{V}\otimes \rho _{V})R(z)\in \mathrm{End}(V\otimes V)$. By
abuse of notation we will refer to $R^{VV}(z)$ henceforth simply by $R(z)$
in order to unburden the formulas.

In addition, we will restrict ourselves to degree zero R-matrices. Recall
that according to the computational rules of supervector spaces operators
carry a degree and are represented by supermatrices whose entries are in
general Grassmann numbers. For the integrable models studied in the
literature one usually assumes that the corresponding $R$-matrix is of
degree zero (see e.g. \cite{GYB}) 
\begin{equation}
\deg R(z):=|a|+|b|+|c|+|d|+\deg R(z)_{ab}^{cd}=0\;.
\end{equation}
and that the non-vanishing matrix elements $R(z)_{ab}^{cd}$ viewed as
Grassmann numbers are also of even degree, $\deg R(z)_{ab}^{cd}=0$. Here the
indices $a,b,c,d=1,...,\dim V$ refer to some homogeneous basis $%
\{v_{a}\}\subset V$ and $|a|=0,1$ denotes the degree of the basis vector $%
v_{a}$. In fact, one wants $R(z)_{ab}^{cd}$ to be ordinary positive numbers
which can be interpreted as Boltzmann weights. Then there exists a
well-defined statistical lattice model whose partition function can be
written as a supertrace over the $L$-fold tensor product 
\begin{equation}
Z(w)=\mathrm{str}_{V^{\otimes L}}T(w)^{M}  \label{Z}
\end{equation}
Here $T(z)$ denotes the graded transfer matrix which is defined as the
partial supertrace of the following operator product 
\begin{equation}
T(z)=\mathrm{str}_{V_{0}}R_{0L}(z)R_{0L-1}(z)\cdots R_{02}(z)R_{01}(z)\in 
\mathrm{End}(V_{1}\otimes V_{2}\cdots \otimes V_{L})  \label{trans}
\end{equation}
The 'auxiliary space' $V_{0}\cong V$ labels the boundary values and the
remaining spaces $V_{1}\otimes V_{2}\cdots \otimes V_{L},V_{i}\cong V$ form
one row of the lattice. The lower indices indicate on which copy of the
representation space the R-matrix acts. It needs to be emphasized that by
the choice (\ref{trans}) of the transfer matrix we have obviously imposed
periodic boundary conditions leading to translation invariance. Assuming as
usual regularity of the R-matrix 
\begin{equation*}
R(z=1)=\pi 
\end{equation*}
with $\pi $ being the previously introduced graded permutation operator (\ref
{pi}) one obtains from the transfer matrix at $z=1$ the translation operator 
\begin{equation}
T(z=1)=\mathrm{str}_{0}R_{0L}(1)R_{0L-1}(1)\cdots R_{01}(1)=\pi _{0L}\cdots
\pi _{02}\pi _{01}=:\Pi ^{-1}\;.
\end{equation}
The action of $\Pi $ amounts to the following shift in one row of the
lattice, 
\begin{equation}
\Pi :V_{1}\otimes V_{2}\cdots \otimes V_{L}\rightarrow V_{2}\otimes
V_{3}\cdots \otimes V_{L}\otimes V_{1}\;.  \label{P}
\end{equation}
That the transfer matrix commutes with the translation operator might be
directly verified from its definition or from the more general formula 
\begin{equation}
\lbrack T(z),T(w)]=0\;.  \label{int}
\end{equation}
That the transfer matrices of different spectral parameters commute is a
direct consequence of the graded Yang-Baxter equation (\ref{gyb}) and
manifests the integrability of the model.

Besides the transfer matrix one is often also interested in the spectrum of
the formally associated spin-chain Hamiltonian which is defined by\footnote{%
Note that whether the spin-chain Hamiltonian is hermitian or not might
depend on the value of the deformation parameter and the chosen
representation $V$.} 
\begin{equation}
H=\left. z\frac{d}{dz}\ln T(z)\right| _{z=1}=\sum_{n=1}^{L}\pi _{nn+1}\left.
z\frac{d}{dz}R_{nn+1}(z)\right| _{z=1}\;,\quad L+1\equiv 1\;.  \label{H}
\end{equation}
Here $\pi _{nn+1}$ denotes the graded permutation operator, acting on the $%
n^{\text{th}}$ and $(n+1)^{\text{th}}$ factor in the spin chain. In the
second equation we have once more exploited the regularity property. From
this expression or the commutation relation one immediately infers that also
the Hamiltonian is translation invariant $[H,\Pi ]=0$.

In contrast, the action of the quantum affine superalgebra $U_{q}(\hat{g})$
on the spin-chain $V_{1}\otimes V_{2}\cdots \otimes V_{L}$ is in general not
translation invariant as can be directly seen from the explicit expression
of its generators in the $L$-fold tensor product 
\begin{eqnarray}
E_{i} &=&\sum_{n=1}^{L}E_{i;n}:=\Delta ^{(L)}(e_{i})=\sum_{n=1}^{L}q^{\frac{%
h_{i}}{2}}\otimes \cdots \otimes q^{\frac{h_{i}}{2}}\otimes e_{i}\otimes q^{-%
\frac{h_{i}}{2}}\otimes \cdots \otimes q^{-\frac{h_{i}}{2}}  \notag \\
F_{i} &=&\sum_{n=1}^{L}F_{i;n}:=\Delta ^{(L)}(f_{i})=\sum_{n=1}^{L}q^{\frac{%
h_{i}}{2}}\otimes \cdots \otimes q^{\frac{h_{i}}{2}}\otimes f_{i}\otimes q^{-%
\frac{h_{i}}{2}}\otimes \cdots \otimes q^{-\frac{h_{i}}{2}}  \notag \\
q^{H_{i}} &=&\prod_{n=1}^{L}q^{H_{i;n}}:=\Delta
^{(L)}(q^{h_{i}})=q^{h_{i}}\otimes \cdots \otimes q^{h_{i}}\quad .  \label{L}
\end{eqnarray}
We stress also that these generators (except for the Cartan elements $%
q^{H_{i}}$) for generic values of the deformation parameter $q$ do not
commute neither with the transfer matrix (\ref{trans}) nor with the
Hamiltonian (\ref{H}) due to the periodic boundary conditions.

\subsection{Translation invariance at roots of unity}

We now demonstrate that at roots of unity certain subalgebras of $U_{q}^{%
\text{res}}(\hat{g})$ as discussed in the preceding section are translation
invariant. For this purpose we state first for generic $q$ the
transformation law for the restricted generators $E_{i}^{(m)},m\in \mathbb{N}
$ , 
\begin{eqnarray}
\Pi E_{i}^{(m)}\Pi ^{-1} &=&E_{i}^{(m)}q^{mH_{i;L}}+  \label{tL} \\
&&\hspace{-0.5in}\hspace{-0.5in} \sum_{n=1}^{m}(-)^{|i|\frac{n(n-1)}{2}%
}q_{i}^{n(m-1)}E_{i}^{\left( m-n\right) }E_{i;L}^{\left( n\right)
}q^{mH_{i;L}}\prod_{l=0}^{n-1}\left(
(-)^{|i|(m+l+1)}q_{i}^{-2l}q^{-H_{i}}-1\right).  \notag
\end{eqnarray}
The proof proceeds by induction and is detailed in the appendix. An
analogous formula holds for $F_{i}^{(m)}$. Letting the deformation parameter
now approach a (primitive) root of unity, $q^{N}\rightarrow 1,$ and setting $%
m=N^{\prime }/2,N^{\prime },2N^{\prime }$ we discuss translation invariance
of the Chevalley-Serre step operators treating as before the cases of $%
\alpha _{i}$ being even or odd separately.

\begin{description}
\item[\textbf{$\boldsymbol{\alpha_{i}}$ white.}]  For white roots ($|i|=0$)
the same considerations as in \cite{KM} apply and we find that the $sl_{2}$
subalgebra commutes or anticommutes with $\Pi $ depending on whether $h_{i}$
takes on even or odd integer values in the chosen highest weight
representation $V$ of the spin chain (see \cite{KM} for details).
Explicitly, the $l=0$ term in the product of formula (\ref{tL}) always
vanishes provided that $q^{H_{i}}=1$ and we obtain 
\begin{equation}
\Pi E_{i}^{(N^{\prime })}\Pi ^{-1}=E_{i}^{(N^{\prime })}q^{N^{\prime
}H_{i;L}}
\end{equation}
The commensurability condition (\ref{commp}) has therefore for even roots of
unity to be strengthened to $\lambda (H_{i})=0\;\mathrm{mod}\;N$.
Furthermore, we conclude that if $q^{N^{\prime }H_{i;L}}=1$ or $-1$ the
generator $E_{i}^{(N^{\prime })}$ commutes or anticommutes with the
translation operator.

\item[\textbf{$\boldsymbol{\alpha_{i}}$ black.}]  For black roots ($|i|=1$)
we discuss the case of even roots of unity with $N^{\prime }$ even first.
Then the generator $E_{i}^{(N^{\prime }/2)}$ is odd and the sign factor in
the product of (\ref{tL}) vanishes. As before we find $\Pi E_{i}^{(\frac{%
N^{\prime }}{2})}\Pi ^{-1}=E_{i}^{(\frac{N^{\prime }}{2})}q^{\frac{N^{\prime
}}{2}H_{i;L}}$ provided that $q^{H_{i}}=1$. However, unlike in the white
root case the sign of $q^{H_{i}}$ alternates in a given multiplet due to the
supercommutation relations (\ref{b1a}). Therefore, the odd generators of
black roots are in general not translation invariant.

The situation looks better for the even generators $E_{i}^{(N^{\prime
})},F_{i}^{(N^{\prime })}$. Now the sign factor $(-)^{|i|(N^{\prime }+l+1)}$
in the product of (\ref{tL}) does not vanish and we have to change the
commensurability condition such that $q^{H_{i}}=-1$ in order to compensate
it. Due to the supercommutation relations (\ref{b2a}) we see that under the
action of the step operators the sign of $q^{H_{i}}$ stays constant in a
given multiplet and from (\ref{tL}) we thus conclude that for $\lambda
(H_{i}/N^{\prime })\in 2\mathbb{Z}+1$ one has 
\begin{equation}
\Pi E_{i}^{(N^{\prime })}\Pi ^{-1}=E_{i}^{(N^{\prime })}q^{N^{\prime
}H_{i;L}}\;\text{.}
\end{equation}
As in the case of white generators the value of the remaining factor $%
q^{N^{\prime }H_{i;L}}$ depends on the chosen representation $V$ the
spin-chain is build out of. If the Cartan generator $h_{i}$ takes on even
integer values in $V$ then the generators commute with the translation
operator, if it is odd integer valued they anticommute.

Similar, one has for even roots of unity $q^{N}=1\,$with $N^{\prime }$ being
odd that in the sector $q^{H_{i}}=-1,\;\lambda (H_{i}/N^{\prime })\in 2%
\mathbb{Z}+1$ the even generator $E_{i}^{(N)}$ satisfies 
\begin{equation*}
\Pi E_{i}^{(N)}\Pi ^{-1}=E_{i}^{(N)}\;.
\end{equation*}
Note that there is this time no additional factor since the generator is
taken to the power $N=2N^{\prime }$. We point out that at odd roots of unity
there is no obvious way to make the generator $E_{i}^{(2N)}$ translation
invariant, since the desired property $q^{H_{i}}=-1$ cannot be satisfied
here.

\item[\textbf{$\boldsymbol{\alpha_{i}}$ grey.}]  Since in this case the step
operators are nilpotent, the only possibility is that the generators $%
E_{i},F_{i}$ are translation invariant. However, this is not the case.
\end{description}

We summarize the results of this section in the following table.

\begin{center}
\begin{tabular}{|c|c|c|c|c|}
\hline\hline
$\alpha $ & $N$ & commensurate sector & generators & invariant \\ 
\hline\hline
white & odd/even & $\lambda (H_{i})=0\;\mathrm{mod}\;N$ & $E_{i}^{(N^{\prime
})},F_{i}^{(N^{\prime })}$ & yes \\ \hline\hline
black & $N^{\prime }$ even & $\lambda (H_{i})=0\;\mathrm{mod}\;N^{\prime }$
& $E_{i}^{(\frac{N^{\prime }}{2})},F_{i}^{(\frac{N^{\prime }}{2})}$ & no \\ 
\hline\hline
&  & $\lambda (H_{i})/N^{\prime }\in 2\mathbb{Z}+1$ & $E_{i}^{(N^{\prime
})},F_{i}^{(N^{\prime })}$ & yes \\ \hline\hline
& $N^{\prime }$ odd & $\lambda (H_{i})/N^{\prime }\in 2\mathbb{Z}+1$ & $%
E_{i}^{(2N^{\prime })},F_{i}^{(2N^{\prime })}$ & yes \\ \hline\hline
& odd & $\lambda (H_{i})=0\;\mathrm{mod}\;N$ & $E_{i}^{(2N)},F_{i}^{(2N)}$ & 
no \\ \hline\hline
grey & odd, even & - & $E_{i},F_{i}$ & no \\ \hline\hline
\end{tabular}
\smallskip
\end{center}

\noindent {\small Table 5.1. Shown are the various step operators which have
a non-trivial root of unity limit and their invariance properties w.r.t. to
the adjoint action of the translation operator.\medskip }

Since according to our analysis the step operators $E_{i}^{(2N^{\prime
})},F_{i}^{(2N^{\prime })}$ associated with a black simple root are only
translation invariant at even roots of unity due to the grading we might in
general not expect that the whole algebraic structure as obtained by the
quantum Frobenius homomorphism of the previous section is compatible with
periodic boundary conditions. Nevertheless, there are several cases where
the entire even subalgebra $\hat{g}_{0}$ is translation invariant (at even
roots of unity), e.g. $\hat{g}=osp(2|1)^{(1)},$ $osp(2|2)^{(2)}$ or $%
sl(1|3)^{(4)}$.

\subsection{Boost operator}

After having established the analogous results for the restricted quantum
superalgebra $U_{q}^{\text{res}}(\hat{g})$ as in the non-graded case we are
now prepared to apply the analogous line of argument as presented in \cite
{KM} to show that the various translation invariant subalgebras listed above
constitute symmetries of the vertex model. For the Hamiltonian (\ref{H}) one
shows directly by exploiting the quantum algebra invariance of the $R$%
-matrix (\ref{jime}) that it commutes with the respective even subalgebras.
In order to show the invariance of the transfer matrix we act with $w\frac{d%
}{dw}$ on the graded Yang-Baxter equation 
\begin{equation*}
R_{0j}(z)R_{0j+1}(w)R_{jj+1}(w/z)=R_{jj+1}(w/z)R_{0j+1}(w)R_{0j}(z)
\end{equation*}
and setting $z=w$ afterwards to find 
\begin{equation*}
\left[ \pi _{jj+1}R_{jj+1}^{\prime }(1),R_{0j+1}(z)R_{0j}(z)\right]
=R_{0j+1}(z)R_{0j}^{\prime }(z)-R_{0j+1}^{\prime }(z)R_{0j}(z)\;.
\end{equation*}
Here the prime indicates acting with $z\tfrac{d}{dz}$ on the respective $R$%
-matrix and we have employed the regularity property $R(1)=\pi $. From this
relation together with translation invariance of the transfer matrix one
finds that the boost operator \cite{boost0,boost1,boost2,boost3} defined as 
\begin{equation}
K=\sum_{j\func{mod}L}j\,\pi _{jj+1}\,\left. z\frac{d}{dz}R_{jj+1}(z)\right|
_{z=1}  \label{boost}
\end{equation}
satisfies the crucial relation 
\begin{equation}
\lbrack K,T(z)]=z\frac{d}{dz}T(z)\;.  \label{boost1}
\end{equation}
Note that the sum is taken over the integers modulo $L$ and that the $R$%
-matrix is of even degree. Upon integration the last relation implies that
under the adjoint action of the boost operator the transfer matrix is
shifted in the spectral parameter, 
\begin{equation}
w^{K}T(z)w^{-K}=T(zw)\;.  \label{boost2}
\end{equation}
Exploiting now repeatedly translation invariance of the generators listed in
Table 5.1, we find first that the boundary terms 
\begin{equation}
\pi _{L1}\,\left. z\frac{d}{dz}R_{L1}(z)\right| _{z=1}
\end{equation}
of the boost operator commute with the respective subalgebras under the
stated commensurability condition. The remaining terms commute also due to
quantum algebra invariance of the $R$-matrix (\ref{jime}) and we conclude
that the boost operator is invariant under the action of the respective
subalgebras. Exploiting translation invariance for the second time we see
from relation (\ref{boost2}) that the transfer matrix commutes with the
translation invariant generators listed in Table 5.1.

For each white and black root we have therefore an $U(sl_{2})$ invariance of
the statistical model in the various commensurate sectors. These different
subalgebras can combine to a larger symmetry algebra. For example when $\hat{%
g}=sl(2m|2n)^{(1)}$ we have as symmetry algebra $sl_{m}\oplus sl_{n}\oplus
u(1)$ at odd roots of unity. The associated class of integrable lattice
theories includes among other the Perk-Schultz models \cite{Perk} and the $U$%
-model \cite{Bris,BKZ}. If the order of the root of unity is even we mention
once more the $osp(2|2)^{(2)}$ example \cite{Kob} whose symmetry algebra is $%
sl_{2}^{(1)}$.

\section{Conclusions}

In this article we have introduced the restricted quantum affine
superalgebra $U_{q}^{\text{res}}(\hat{g})$ and investigated its properties
in the root of unity limit. We have proved several new identities (for
arbitrary values of the deformation parameter $q$) such as the
supercommutator (\ref{F1}) or the higher order Serre relations for quantum
affine algebras (\ref{HOS}). In the root of unity limit $q^{N}\rightarrow 1$
we then employed these new formulas to prove an analogue of Lustzig's
quantum Frobenius homomorphism for the super case. While the odd or
fermionic part of the superalgebra does not give rise to interesting
structures in this limit, we showed that for the even part $U_{q}^{\text{res}%
}(\hat{g})_{0}$ one recovers the ''classical'' algebraic relations. Namely,
for superalgebras whose distinguished simple root system contains white and
black roots only one obtains the entire even non-deformed subalgebra $U(\hat{%
g}_{0})$ as $q^{N}\rightarrow 1$ with $N$ being odd. When grey roots are
present we had to restrict ourselves to a subalgebra $\hat{g}_{0}^{\text{%
trunc}}\subseteq \hat{g}_{0}^{{}}$. As we pointed out in the text a further
extension of the theorem might be possible by investigating suitably defined 
$q$-deformed Cartan-Weyl bases in the root of unity limit. Since their
structure, especially for higher tensor products, is more involved we will
leave this issue to future work.

Another point which deserves investigation is the finite-dimensional
representation theory of the restricted quantum superalgebra at roots of
unity. While for the non-graded case the representations of $U_{q}^{\text{res%
}}(\hat{g})$ have been classified in \cite{Lus} and \cite{CP2} similar
results do not exist in the literature for the super case. Our discussion of
the restricted quantum superalgebra $U_{q}^{\text{res}}(\hat{g})$ is a first
step in this direction.

Applying the results of our mathematical discussion we have demonstrated
that the non-deformed even subalgebra $U(\hat{g}_{0})$ obtained from $U_{q}^{%
\text{res}}(\hat{g})$ at roots of unity either equals or contains proper
non-abelian symmetry algebras of integrable lattice models. For instance, we
established that for the special linear supergroups $\hat{g}=sl(m|n)^{(1)}$
the symmetry algebra is $sl_{m}\oplus sl_{n}\oplus u(1)$ at odd roots of
unity. The associated class of integrable lattice theories includes among
others the Perk-Schultz models \cite{Perk} and the $U$-model \cite{Bris,BKZ}
which is a generalization of the Hubbard model with correlated hopping terms
in the associated spin-chain Hamiltonian. An example involving black roots
only are the models based on $osp(2|2)^{(2)}$ (see e.g. \cite{Kob}) whose
symmetry algebra we identified to be $\hat{g}_{0}=sl_{2}^{(1)}$.

A further step is to relate the finite-dimensional representation theory of
the restricted quantum groups and the symmetry algebras to the Bethe ansatz.
Since a representation independent formulation of the Bethe ansatz is not
known, this can be done by choosing one of the earlier mentioned physical
models, which are formulated in a specific representation of the quantum
affine superalgebra.\vspace{0.2cm}\\
\\
{\bf Acknowledgement.} 
We would like to thank A. Foerster, P.
Kulish, J. Links for valuable comments. We are also grateful to B.M. McCoy for 
stimulating and encouraging discussions as well as helpful comments on a draft 
version of this paper. CK is financially supported by the NSF grants DMR-0073058 
and PHY-9988566 and IR by CNPq (Conselho Nacional de Desenvolvimento
 Cient\'{\i}fico e Tecnol\'{o}gico) and PRONEX.\\

\pagebreak \appendix

\section{Proofs}

\subsection{Supercommutator of the restricted algebra}

The formula we like to prove for $m\geq n$ reads 
\begin{equation*}
\lbrack e_{i}^{m},f_{i}^{n}]=\sum_{k=1}^{n}(-)^{|i|(m-k)(n-k)}\qatopd{m}{k}%
_{q_{i}}\qatopd{n}{k}_{q_{i}}\left[ k\right]
_{q_{i}}!f_{i}^{n-k}e_{i}^{m-k}\prod_{l=1}^{k}\left[ h_{i};m-n-l+1\right] \;.
\end{equation*}
Induction start. 
\begin{equation*}
\lbrack e_{i}^{m},f_{i}]=[m]_{q_{i}}e_{i}^{m-1}[h_{i};m-1]
\end{equation*}
Induction step. 
\begin{eqnarray*}
\lbrack e_{i}^{m},f_{i}^{n+1}]
&=&(-)^{|i|m}f_{i}[e_{i}^{m},f_{i}^{n}]+[e_{i}^{m},f_{i}]f_{i}^{n} \\
&=&(-)^{|i|m}f_{i}[e_{i}^{m},f_{i}^{n}]+[m]_{q_{i}}e_{i}^{m-1}f_{i}^{n}[h_{i};m-2n-1]
\end{eqnarray*}
\begin{eqnarray*}
\lbrack e_{i}^{m},f_{i}^{n+1}] &=&\sum_{k=1}^{n}(-)^{|i|(x_{k}(m,n)+m)}%
\qatopd{m}{k}_{q_{i}}\qatopd{n}{k}_{q_{i}}\left[ k\right]
_{q_{i}}!f_{i}^{n+1-k}e_{i}^{m-k}\prod_{l=0}^{k-1}\left[ h_{i};m-n-l\right]
\\
&&+\sum_{k=2}^{n+1}(-)^{|i|x_{k-1}(m-1,n)}\qatopd{m}{k}_{q_{i}}\qatopd{%
n+1}{k}_{q_{i}}\left[ k\right] _{q_{i}}!\frac{[k]_{q_{i}}}{[n+1]_{q_{i}}}%
\,f_{i}^{n+1-k}e_{i}^{m-k} \\
&&\times \lbrack h_{i};m-2n-1]\prod_{l=1}^{k-1}\left[ h_{i};m-n-l\right] \\
&&\quad +(-)^{|i|(m-1)n}[m]_{q_{i}}f_{i}^{n}e_{i}^{m-1}[h_{i};m-2n-1]
\end{eqnarray*}
Taking into account that $x_{k}(m,n)=(m-k)(n-k)\;\mathrm{mod}\;2$ one
verifies the following identities, 
\begin{multline*}
\lbrack n+1]_{q_{i}}(-)^{|i|(x_{k}(m,n+1)-m)}[h_{i};m-n-k]= \\
(-)^{|i|x_{k}(m,n)}[n+1-k]_{q_{i}}[h_{i};m-n]+(-)^{|i|(x_{k-1}(m-1,n)+m)}[k]_{q_{i}} 
\left[ h_{i};m-2n-1\right]
\end{multline*}
\begin{multline*}
(-)^{|i|x_{1}(m,n+1)}[n+1]_{q_{i}}[h_{i};m-n-1]= \\
(-)^{|i|x_{1}(m,n)}[n]_{q_{i}}[h_{i};m-n]+(-)^{|i|(m-1)n}[h_{i};m-2n-1]
\end{multline*}
\begin{equation*}
(-)^{|i|x_{n+1}(m,n+1)}=(-)^{|i|x_{n}(m-1,n)}\;.
\end{equation*}
From these equations the desired formula for $n\rightarrow n+1$ follows,
which completes the induction proof.

\subsection{Proof of formula (\ref{def})}

Defining $\Theta _{m,n}^{\pm }$ by means the $q$-deformed adjoint action 
\begin{equation*}
\Theta _{m,n}^{\pm }:=\frac{\left( \mathrm{ad}_{q^{\pm 1}}e_{i}\right)
^{m}e_{j}^{n}}{[m]_{q_{i}^{\pm 1}}![n]_{q_{i}^{\pm 1}}!}
\end{equation*}
we are going to prove the identity 
\begin{equation*}
\Theta _{m,n}^{\pm }=\sum_{r+s=m}(-)^{s+x_{s}^{n}}q_{i}^{\mp
s(1-na_{ij}-m)}e_{i}^{(r)\pm }e_{j}^{(n)\pm }e_{i}^{(s)\pm }
\end{equation*}
with the degree function equal to 
\begin{equation*}
x_{s}^{n}:=|i|\frac{s(s-1)}{2}+|i||j|ns\;.
\end{equation*}
Here the $\pm $ sign in the upper index of the restricted step operators
refers to $q^{\pm 1}$. In the following we set $y:=|i|m+|i||j|n$. One then
verifies by means of equation (\ref{2}) that 
\begin{eqnarray*}
\lbrack e_{i},\Theta _{m,n}^{\pm }]_{q^{\pm 1}} &=&e_{i}\Theta _{m,n}^{\pm
}-(-)^{y}q_{i}^{\pm (na_{ij}+2m)}\Theta _{m,n}^{\pm }e_{i} \\
&=&\sum_{r+s=m+1}(-)^{s+x_{s}}q_{i}^{\mp s(1-na_{ij}-m)}e_{i}^{(r)\pm
}e_{j}^{(n)\pm }e_{i}^{(s)\pm }[r]_{q_{i}^{\pm 1}} \\
&&+\sum_{r+s=m+1}(-)^{s+x_{s-1}+y}q_{i}^{\mp s(1-na_{ij}-m)}q_{i}^{\pm
(m+1)}e_{i}^{(r)\pm }e_{j}^{(n)\pm }e_{i}^{(s)\pm }[s]_{q_{i}^{\pm 1}} \\
&=&\sum_{r+s=m+1}(-)^{s+x_{s}}q_{i}^{\mp s(-na_{ij}-m)}e_{i}^{(r)\pm
}e_{j}^{(n)\pm }e_{i}^{(s)\pm }\left\{ q_{i}^{\mp s}[r]_{q_{i}^{\pm
1}}\right. \\
&&\left. +(-)^{x_{s-1}+x_{s}+y}q_{i}^{\pm r}[r]_{q_{i}^{\pm 1}}\right\} \\
&=&[m+1]_{q_{i}^{\pm 1}}\Theta _{m+1,n}^{\pm }\;.
\end{eqnarray*}

\subsection{Proof of the translation formula}

We state the proof for $E_{i}^{(m)}$ only, the one for $F_{i}^{(m)}$ being
completely analogous. For generic $q$ one finds the following relations 
\begin{equation*}
\Pi \,E_{i}\,\Pi ^{-1}=E_{i}\,q^{H_{i;L}}+E_{i;L}(q^{-H_{i}}-1)q^{H_{i;L}}\;,
\end{equation*}
where use has been made of the straightforward identities 
\begin{eqnarray*}
\Pi \,E_{i;n}\,\Pi ^{-1} &=&E_{i;n-1}q^{H_{i;L}}\quad n>1 \\
\Pi \,q^{H_{i;n}}\,\Pi ^{-1} &=&q^{H_{i;n-1}}
\end{eqnarray*}
We claim that the transformation property for the $m^{\text{th}}$ power
reads 
\begin{multline*}
\Pi E_{i}^{m}\,\Pi ^{-1}=\sum_{n=0}^{m}(-)^{|i|\frac{n(n-1)}{2}%
}q_{i}^{n(m-1)}\qatopd{m}{n}_{q_{i}}E_{i}^{m-n}E_{i;L}^{n}q^{mH_{i;L}}%
\times \\
\prod_{l=0}^{n-1}\left( (-)^{|i|(m+l+1)}q_{i}^{-2l}q^{-H_{i}}-1\right) \;.
\end{multline*}
Proceeding by induction we assume that the above relation holds for $m$ and
calculate 
\begin{multline*}
\Pi E_{i}^{m+1}\Pi ^{-1}=\sum_{n=0}^{m}(-)^{|i|\frac{n(n-1)}{2}%
}E_{i}^{m-n}E_{i;L}^{n}E_{i}\,q_{i}^{n(m-1)}\qatopd{m}{n}%
_{q_i}q^{(m+1)H_{i;L}}\times \\
\prod_{l=1}^{n}\left( (-)^{|i|(m+l)}q_{i}^{-2l}q^{-H_{i}}-1\right) \\
+\sum_{n=0}^{m}(-)^{|i|\frac{n(n-1)}{2}}E_{i}^{m-n}E_{i;L}^{n+1}%
\,q_{i}^{n(m-1)}\qatopd{m}{n}_{q_i}\left( q_{i}^{2m}q^{-H_{i}}-1\right)
q^{(m+1)H_{i;L}}\times \\
\prod_{l=1}^{n}\left( (-)^{|i|(m+l)}q_{i}^{-2l}q^{-H_{i}}-1\right)
\end{multline*}
Employing the commutation relations 
\begin{equation*}
E_{i;L}^{n}E_{i}=(-)^{|i|n}q_{i}^{2n}E_{i}E_{i;L}^{n}+E_{i;L}^{n+1}(1-(-)^{|i|n}q^{2n})
\end{equation*}
one derives 
\begin{multline*}
\Pi E_{i}^{m+1}\Pi ^{-1}=E_{i}^{m+1}q^{(m+1)H_{i;L}}+ \\
(-)^{|i|\frac{(m+1)m}{2}%
}E_{i;L}^{m+1}q_{i}^{m(m-1)}q_{i}^{2m}q^{(m+1)H_{i;L}}%
\prod_{l=0}^{m}((-)^{|i|(m+l)}q_{i}^{-2l}q^{-H_{i}}-1)+ \\
\sum_{n=1}^{m}(-)^{|i|\frac{n(n-1)}{2}}E_{i}^{m+1-n}E_{i;L}^{n}\,\frac{%
q_{i}^{nm}q^{(m+1)H_{i;L}}}{[m+1]_{q_{i}}}\qatopd{m+1}{n}_{q_{i}}\left\{
...\right\}\times \\
\prod_{l=1}^{n-1}\left( (-)^{|i|(m+l)}q_{i}^{-2l}q^{-H_{i}}-1\right)
\end{multline*}
where the term in the brackets reads 
\begin{eqnarray*}
\{...\}
&=&[m+1-n]_{q_{i}}(-)^{|i|m}q_{i}^{-n}q^{-H_{i}}+[n]_{q_{i}}(-)^{|i|(1-n)}q_{i}^{m+1-n}q^{-H_{i}}
\\
&&\quad \quad -[m+1-n]_{q_{i}}(-)^{|i|n}q_{i}^{n}-[n]_{q_{i}}q_{i}^{-m-1+n}
\\
&=&[m+1]_{q_{i}}((-)^{|i|m}q^{-H_{i}}-1)\;.
\end{eqnarray*}
This completes the proof. Note that in the last step use has been made of
the elementary relation (\ref{2}) for $q$-deformed integers.

\end{document}